\documentclass[mathleft
]{an}
\usepackage{graphicx}
\usepackage{times}
\usepackage{natbib}
\begin{document}
\sloppy

\Pagespan{0}{}
\Yearpublication{}%
\Yearsubmission{2010}%
\Month{03}%
\Volume{}%
\Issue{}%

\title{Magneto-acoustic wave propagation and mode conversion in a
  magnetic solar atmosphere: comparing results from the CO$^5$BOLD
  code with ray theory.}

\author{C.~Nutto\thanks{Corresponding author:
    \email{nutto@kis.uni-freiburg.de}} \and O.~Steiner\and
  M.~Roth}

\titlerunning{Magneto-acoustic wave propagation}
\authorrunning{C. Nutto, O. Steiner \& M. Roth}
\institute{
Kiepenheuer-Institut f\"ur Sonnenphysik, Sch\"oneckstr. 6, 79104 Freiburg, Germany}

\received{2010 March 31}
\accepted{2010 August 6}
\publonline{}

\abstract{%
  We present simulations of magneto-acoustic wave propagation in a
  magnetic, plane-parallel stratified solar model atmosphere,
  employing the CO$^5$BOLD-code. The tests are carried out for two
  models of the solar atmosphere, which are similar to the ones used
  by \citet{nutto_cally07} and \citet{nutto_schunker06}. The two
  models differ only in the orientation of the magnetic field. A
  qualitative comparison shows good agreement between the numerical
  results and the results from ray theory. The tests are done in view
  of the application of the present numerical code for the computation
  of energy fluxes of propagating acoustic waves into a dynamically
  evolving magnetic solar atmosphere. For this, we consider waves with
  frequencies above the acoustic cut-off frequency.}


\maketitle

\section{Introduction}
Investigations by \citet{nutto_rosenthal02} and \citet{nutto_bogdan03}
have shown that the concept of mode conversion is crucial for
understanding the propagation of magneto-acoustic waves in the
magnetic solar atmosphere. Thus, various MHD-codes have been developed
to study linear and non-linear magneto-acoustic wave propagation in
stellar atmospheres. This is done either by linearization of the
equations around an initial atmospheric configuration
\citep{nutto_cameron08} or by solving the non-linear equation for the
deviations from the initial state of the atmosphere. The latter
approach was used by \citet{nutto_khomenko06} and by
\citet{nutto_shelyag09} for the study of wave propagation within and
in the environment of a sunspot.  \citet{nutto_steiner07} have
considered wave propagation in a convectively unstable atmosphere that
contained a network magnetic element, which was part of the dynamical
evolution. For this they did not compute the deviation from an initial
state: instead they computed the evolution of the entire,
nonstationary stratification twice, once with the wave generating
perturbation and once without it. The wave was made visible by
subtraction of the numerical solution without the perturbation from
that including the perturbation. This method has the advantage that it
allows us to use a previously existing, well tested code that was
specifically developed for the simulation of dynamical
magneto-convective processes including a realistic equation of state
and radiative transfer with realistic opacities without the need of
major changes. It also enables us to follow the wave propagation in a
dynamically evolving atmosphere and not only as a deviation from an
initial, static state. However this method has the disadvantage to be
prone to numerical inaccuracies. In fact, the results by
\citet{nutto_steiner07} show spurious velocities in the region of
strong magnetic fields (low plasma-$\beta$).

Meanwhile, we have improved the accuracy of the method and the code
used by \citet{nutto_steiner07}. In this paper we carry out a few
calculations for testing the correct behavior of magneto-acoustic wave
propagation and mode conversion. This is necessary as is not obvious
in advance that a numerical scheme for the solution of the system of
MHD equations would correctly reproduce the process of mode
conversion.  To this aim we use the solutions presented by
\citet{nutto_schunker06} and \citet{nutto_cally07} as a test case for
our MHD-code. Using a magnetically modified version of the solar Model
S of \citet{nutto_dalsgaard96}, they studied the wave propagation and
mode conversion analytically as well as numerically.

\section{Numerical setup}\label{nutto_Sect:Method}
\begin{figure}
\centering
\includegraphics[width = 0.35\textwidth]{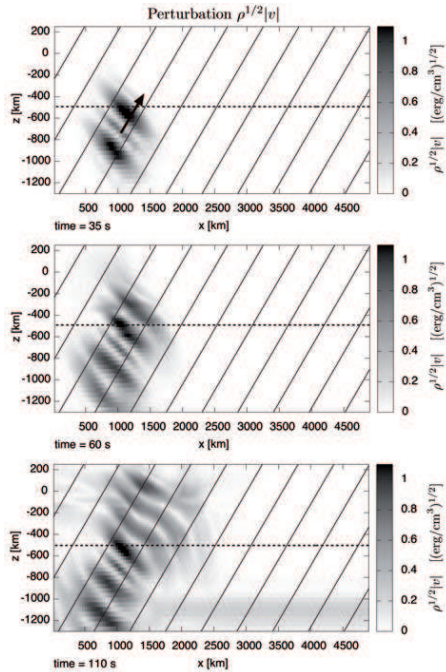} 
\caption{Velocity perturbation scaled with the density, $\rho^{1/2}
  |v|$, for different time steps. The magneto-acoustic wave propagates
  from the source (piston) located at
  $(x,z)=(1000~\mathrm{km},-750~\mathrm{km})$. The position of the
  piston and the direction of the excited wave front is indicated by
  the arrow in the top panel.  The horizontal dashed line shows the
  equipartition layer, where the sound speed equals the Alfv\'en
  speed, $c_{\mathrm{s}}=c_{\mathrm{a}}$. The tilted vertical lines
  represent magnetic field lines.}
\label{nutto_Fig:dvrho_dp_B3000G30Deg}
\end{figure}
The simulations are carried out using the CO$^5$BOLD-code
\citep{nutto_freytag02,nutto_schaffenberger05,nutto_schaffenberger06},
which solves the magnetohydrodynamic equations for a fully
compressible gas including radiative transfer and a realistic equation
of state.\newline The model atmosphere corresponds to the top part of
Model S \citep{nutto_dalsgaard96} interpolated onto the grid of the
simulation domain. Superimposed on the model atmosphere is a
homogenous magnetic field with a field strength of 3000~G. In the
first case of our investigations the magnetic field is tilted by an
angle $\alpha=$~30$^\circ$ to the vertical axis. In the second case
the orientation of the magnetic field with respect to the vertical
axis is $\alpha=$~-30$^\circ$. The atmosphere is then evolved for some
time without applying any perturbations to settle to the
quasi-hydrostatic state compatible with the present
discretization.\newline The vertical extent of the simulation domain
covers a range of 1700~km, of which 1300~km are below and 400~km are
above optical depth unity. The horizontal extent of the atmosphere is
4900~km. In the vertical direction an adaptive grid with 58 cells is
used. The grid resolution at the bottom of the simulation domain is
46~km while the resolution for the photosphere is 7~km. The horizontal
direction is covered by 122 cells with a constant resolution of
40~km. \newline Transmitting boundary conditions are applied in the
lateral and in the vertical direction. The magnetic field is kept
constant at a specific angle to the vertical. This ensures that the
magnetic field keeps its given orientation. Magneto-acoustic waves are
excited using a piston, which is implemented as a small, gaussian
shaped perturbation to the pressure gradient in the momentum
equation. The center of the piston is placed at
$(x,z)=(\mathrm{1000~km},-\mathrm{750~km})$ and the full width at half
maximum of the gaussian is 200~km. The piston continuously launches
sinusoidal waves with a monochromatic frequency of
$f_0=20$~mHz. Although, lower frequencies would be more practical for
actual detections in the solar case, the relatively high frequency was
chosen for a good visualization of the mode transmission and
conversion. Here, we are interested in propagating waves and not in
frequencies near or below the cut-off frequency. One can expect a
similar behavior of the wave propagation for all frequencies above the
acoustic cut-off frequency, according to \citet{nutto_khomenko06} and
references therein, so that the 20~mHz waves should be a good
representation for all waves above $\approx 5$~mHz. \newline The
initial wave vector has an angle of 30$^\circ$ to the vertical
direction. Thus, for the first case the initial wave vector is
parallel to the orientation of the magnetic field vector while in the
second case the angle of incidence of the wave vector to the magnetic
field vector (the ``attack angle'') is about 60$^\circ$. The piston is
placed at a depth where plasma-$\beta \gg 1$. Thus, the excited waves
are predominantly acoustic. \newline In the next section we will show
how the magnetic field affects the propagation of magneto-acoustic
waves.

\section{Wave propagation in inclined magnetic fields}

\subsection{Small attack angle}
First we consider the case where the magnetic field in the solar
atmosphere is inclined by 30$^\circ$ to the vertical direction. Thus,
the movement of the piston is parallel to the magnetic field. This
setup corresponds to the investigated case presented in the top panel
of Fig.~3 in \citet{nutto_cally07} except that we use a stronger
magnetic field and high frequency waves.
\begin{figure*}
\centering
\includegraphics[width = 0.6\textwidth]{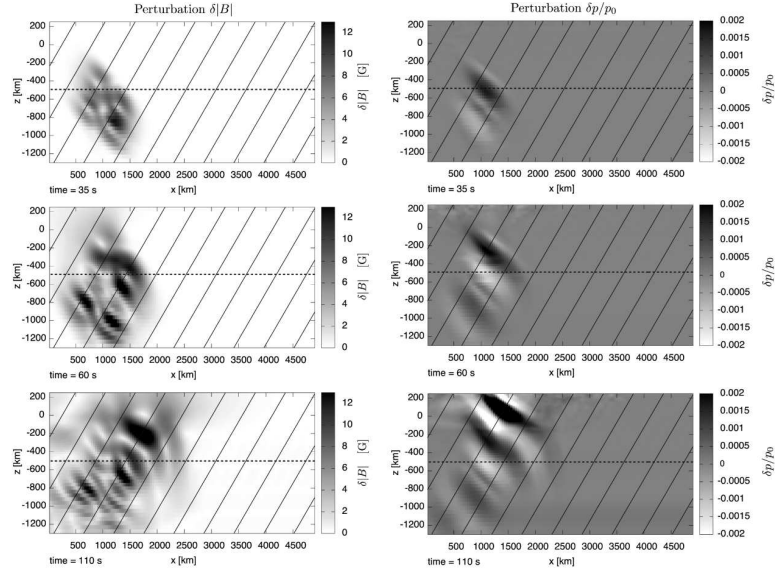} 
\caption{Left column: perturbation $\delta |B|$ of the
  magnetic field strength. Right column: pressure perturbation, $\delta
  p/p_0$, normalized to the local pressure. The horizontal line represents the
  equipartition layer and the tilted lines represent the magnetic
  field lines.}
\label{nutto_Fig:dB_B3000G30Deg}
\end{figure*}
\newline Fig.~\ref{nutto_Fig:dvrho_dp_B3000G30Deg} shows the wave
propagation of the magneto-acoustic wave for the velocity scaled with
the density for different snapshots. In velocity, both
magneto-acoustic waves, the slow and the fast mode, can be followed
since both cause velocity perturbations. However, the different modes
are not easily distinguishable. In order to follow the wave
propagation of the slow mode only, which is predominantly an acoustic
wave, the perturbation of the gas pressure is plotted in the right
column of Fig.~\ref{nutto_Fig:dB_B3000G30Deg}. As mentioned in
Sect.~\ref{nutto_Sect:Method}, the wave is launched by the piston as a
predominantly acoustic wave. Upon reaching the equipartition layer
where the Alfv{\'e}n speed reaches the sound speed, $c_a=c_s$, the
acoustic wave is mostly transmitted, however it changes from the fast
to the slow branch\footnote{The zone where wave modes can undergo
  conversion is a layer of finite thickness around the equipartition
  level. See \citet{nutto_cally07}.}. The slow mode, which is
predominantly an acoustic wave, is then guided along the field lines
into the higher layers of the atmosphere as can be seen in the
pressure perturbation. The evolution of the magnetic field
perturbation is plotted in the left column of
Fig.~\ref{nutto_Fig:dB_B3000G30Deg}. Although the wave is launched
parallel to the magnetic field vector, because of the gradient of the
sound speed in the convection zone the wave vector is tilted gradually
in more vertical direction as the wave travels towards the
photosphere. Hence, at the equipartition layer the wave vector and
magnetic field vector comprise a small angle, allowing only a small
amount of the wave energy to be converted into a fast magneto-acoustic
wave. This fast mode then gets turned around at the bottom of the
photosphere because of the high gradient in Alfv\'en speed. However,
most of the energy remains in the acoustic wave, which then reaches
the higher layers of the atmosphere.

\subsection{Large attack angle}
\begin{figure*}
\centering
\includegraphics[width
=.65\textwidth]{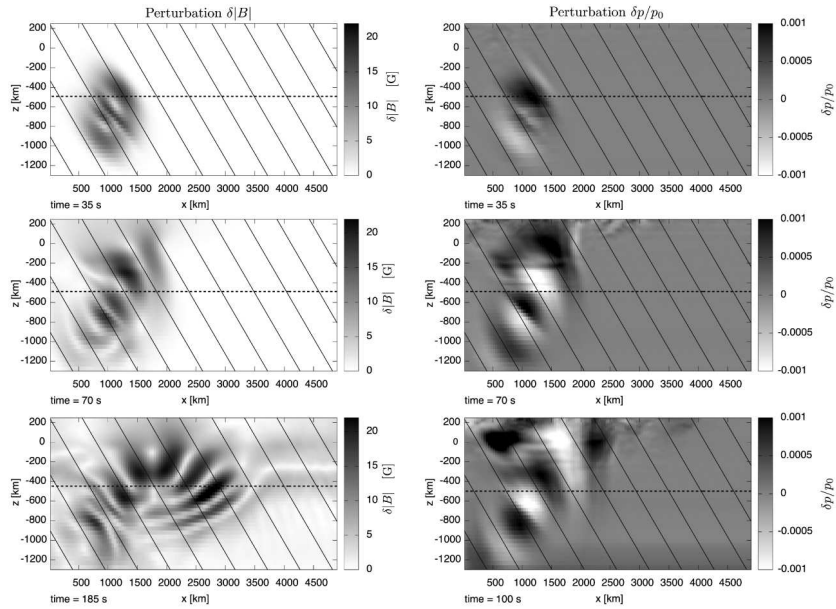}
\caption{Left column: perturbation of the magnetic field strength,
  $\delta |B|$. Right column: perturbation of the pressure, $\delta
  p/p_0$, for different time steps. Note, that the plots in the last
  row can not be compared directly, since the plots show different
  snapshots (left plot at $t = 185$~s; right plot at $t = 100$~s).
  The last snapshot for $\delta |B|$ shows that the wave has undergone
  complete refraction and travels back to the solar interior.}
\label{nutto_Fig:dp_dB_B3000G-30Deg}
\end{figure*}

In the second case, which we consider here, the magnetic field is
inclined by -30$^\circ$ to the vertical direction. The wave vector of
the incident wave however stays the same as in the first case. Thus,
the wave field comprises a large attack angle with the magnetic
field. This case corresponds to the bottom panel of Fig.~3 in
\citet{nutto_cally07}. Fig.~\ref{nutto_Fig:dp_dB_B3000G-30Deg} shows
the perturbation of the pressure and the perturbation of the magnetic
field.
At the equipartition layer the wave front again splits into two parts,
whereby this time the fast magneto-acoustic wave receives more
energy. The wave front that consists of the slow magneto-acoustic wave
is again guided along the field lines into the higher solar
atmosphere, as is visible from the pressure perturbation plotted in
the right column of Fig.~\ref{nutto_Fig:dp_dB_B3000G-30Deg}. The
second and more dominant wave front is the fast magneto-acoustic wave
emerging from the equipartition layer. Again, because of the high
gradient of the Alfv\'en speed, the wave is refracted and turns around
towards the solar interior.

\subsection{Comparison of the acoustic flux for both cases}
\begin{figure}
\centering
\includegraphics[width = .4\textwidth]{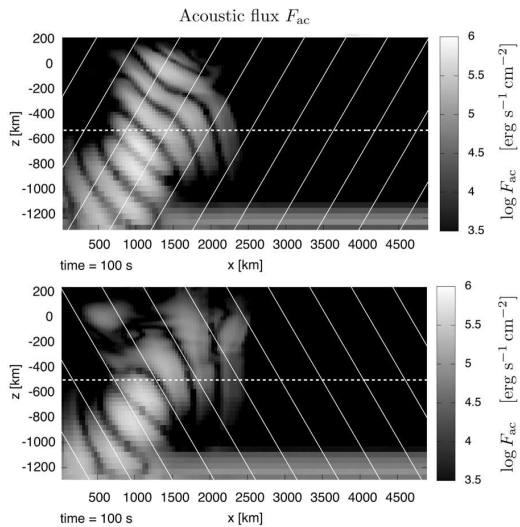}
\caption{Vertical acoustic flux, $F_{\mathrm{ac,z}}$, for both test cases. Top:
  acoustic flux for the atmosphere with the wave vector almost
  parallel to the magnetic field. Bottom: acoustic flux for the case
  of the large attack angle.}
\label{nutto_Fig:fluxes}
\end{figure}

In order to demonstrate how mode conversion is affected by the attack
angle, the vertical acoustic wave flux $F_{\mathrm{ac,z}}= \delta
p*v_z$ is calculated.  A comparison of the vertical acoustic flux for
both cases, small and large attack angle, is plotted in
Fig.~\ref{nutto_Fig:fluxes} for the same time at $t = 100$~s. When the
angle between the wave vector and the magnetic field vector is small,
most of the energy of the wave remains acoustic.  The lower panel
shows the case with the large attack angle. Only little acoustic
energy is transmitted by the slow magneto-acoustic wave, visible in
the upper left hand corner of the lower plot.  The refracted acoustic
flux is due to the nature of the magneto-acoustic wave, which causes
non-vanishing gas pressure perturbations for both the fast and the
slow modes. The fast mode however, is primarily dominated by the
magnetic pressure. The acoustic flux is less than for the case where
incident waves propagate parallel to the magnetic field.\newline Thus,
most of the acoustic flux is transmitted when the wave vector and
magnetic field vector are parallel at the equipartition layer. A large
attack angle causes a major fraction of the wave energy to convert
from acoustic to magnetic. Hence, it is not transported into the
higher layers of the solar atmosphere because of the refraction of the
fast mode.

\section{Conclusions}
Our results compare favorably with the wave propagation path and mode
conversion as derived from ray conversion theory
\citep{nutto_schunker06,nutto_cally07} using a numerical code, which
solves the magnetohydrodynamic equations for an ideal, compressible
medium without any special precautions for the treatment of mode
conversion. This gives us some confidence to apply the present code to
more complex atmospheric configurations. In particular, the present
method can be used to track waves into a dynamically evolving,
convectively unstable atmosphere.\newline The amount of acoustic
energy that is transmitted into the solar atmosphere depends strongly
on the angle between the local wave vector and the magnetic field
vector at the equipartition layer. Thus, this specific layer in the
atmosphere acts as a filter where the slow and fast magneto-acoustic
waves are separated and only the acoustic flux of the slow
magneto-acoustic wave reaches the higher layers of the atmosphere.
\newline \citet{nutto_schunker06} and \citet{nutto_cally07} also test
the wave propagation for vertical magnetic fields. Due to the limited
space, we omit these results here, but note that their and our results
qualitatively agree.

\acknowledgements The authors acknowledge support from the European
Helio- and Astroseismology Network (HELAS), which is funded as
Coordination Action by the European Commission's Sixth Framework
Programme. The authors thank the anonymous referee for the thorough
reading of the manuscript and the constructive comments.

\bibliographystyle{aa}

\end{document}